\documentclass[referee]{raa}            

\usepackage{graphicx,times}             
\usepackage[utf8]{inputenc}
\usepackage[T1]{fontenc}
\usepackage{natbib}
\usepackage{amssymb,amsmath}
\bibpunct{(}{)}{;}{a}{}{,}

\usepackage[pagebackref=true]{hyperref}

\usepackage{changes} 

\begin{document}

  \title{The characteristics of variability of AGNs based on the structure function
}

   \volnopage{Vol.0 (20xx) No.0, 000--000}      
   \setcounter{page}{1}          

   \author{Xuan Wei 
      \inst{1}
   \and Jie Tang
      \inst{1,*}
   \and Yu Tao
      \inst{1}
   }

   \institute{1 School of Physics and Telecommunication Engineering, Shaanxi University of Technology, Hanzhong, 723000, China; {\it tj168@163.com}\\
\vs\no
   {\small }}
\abstract{ Variability is one of the classic features of active galactic nuclei (AGNs). The normalized structure function was applied to distinguish variability samples from OVRO, ASAS-SN and Fermi. A power-law function model was selected to fit the structure functions of samples of three bands. We present the available samples of three bands, and by integrating two parameters, we obtain ideal discrimination results for three bands. Meanwhile, the differences between BL Lacs and FSRQs of Fermi and non-Fermi samples are well verified. The results show that the improved structure function can effectively distinguish samples of radio, optical, and gamma-ray. Additionally, BL Lacs and FSRQs in both Fermi and non-Fermi samples can be distinguished. The conclusion obtained through the distinction of structural functions in different bands supports that the variability in the three bands are caused by different physical mechanisms respectively: the samples in the optical band are radio quiet AGNs, and their variability is mainly caused by the fluctuations of the accretion disk, and the samples of radio band and gamma-ray band are radio loud AGNs whose variability is mainly caused by relativistic jet radiation. This conclusion conforms to the unified standard interpretation of variability about AGNs. Using these two parameters, we verify that there is no fundamental difference between Fermi and non-Fermi BL Lacs, while significant differences exist between FSRQs. However, the power exponent of the two can well distinguish BL Lacs.
}

   \authorrunning{Xuan Wei, Jie Tang \& Yu Tao }            
   \titlerunning{The characteristics of variability of AGNs}  

   \maketitle

%
%
\section{Introduction}           
\label{sect:intro}

Active galactic nucleus (AGNs), a special class of external galaxy, are the cores of an extremely bright active galaxy (\citealt{MaWu2023}). AGNs exist with violent physical processes and phenomena (\citealt{Huang+2005}).We can observe many astronomical phenomena about AGNs because of their violent activity, while variability is one of the most remarkable observation characteristics (\citealt{Yang2022, HuZhang2007}). AGNs can be divided into many subtypes according to their different characteristics, and these subtypes also have variability. The radiation from AGNs covers almost the entire electromagnetic bands, including from the low-frequency radio band to the high-energy gamma ray, and variability can be observed at all bands. AGNs can be divided into radio quiet (RQ) and radio loud (RL) according to the radio loudness ($ R_{ro} = f_{5\mathrm{GHz}} / f_{4400\mathring{\mathrm{A}}} $
) (\citealt{Kellermann+etal+1989}).RQ AGNs include quasars (QSOs) and Seyfert galaxies (Seyfert), while RL AGNs include blazars, the most extreme subclass of AGNs, which can be divided into BL Lacertae objects (BL Lacs) and flat spectrum radio quasars (FSRQs) according to the difference from emission-line features (\citealt{Pei+etal+2020, Su+etal+2024}). Both subcategories are often regarded as research objects in the study of AGNs (\citealt{Hu2006}). \citet{Padovani2017} found that 5 to 10 percent of QSOs in RQ AGNs are FSRQs in RL AGNs. About 30 of the more than 500 high redshift QSOs discovered were classified as RL QSOs by 2004 (\citealt{Romani+etal+2004, McGreer+etal+2006, Zeimann+etal+2011, Sbarrato+etal+2012, Banados+etal+2018}).

Many scholars have studied the variability of AGNs with the development of observational methods and the extension of database. It has been found that some AGNs have period of light variation which can be analyzed by methods such as structure function, power spectrum and Jurkevich. Cross-wavelet analysis and wavelet power spectrum have been used to analyze the main oscillation period and correlation of OJ 287 (\citealt{Tang+2014}). \citet{Sukharev2015} and \citet{KudryavtsevaPyatunina2006} respectively used wavelet analysis and Jurkevich to study the period of light variation about many radio bands of 3C 446. Some studies have found that in addition to the period of light variation, variability also has characteristics of chaotic, fractal and trend (\citealt{Tang+2014}). In the process of the existing research, some scholars believe that variability is a random process (\citealt{Kelly+etal+2009, MacLeod2010, Suberlak+etal+2021, Guo2017, Caplar+etal+2017, Mushotzky+etal+2011, Shen+Burke+2021, Minev+etal+2021, Goyal+etal+2021}), which can be fitted by appropriate models. \citet{MacLeod2010} believed that variability of QSOs could be fitted by the random walk model (DRW), but \citet{Guo2017} and \citet{Kasliwal2015} thought that not all the variability of QSOs could be fitted by DRW.In the current theoretical models explaining the variability mechanism, the model of disk instability also regards the variability as a random event. The accretion disk interacts with the external environment through random factors and energy consumption, leading the violent variability (\citealt{Kawaguchi1998, Harko+etal+2014}). Regarding the possible characteristic of chaotic, the principle of variability can be explained by the nonlinear mechanism (\citealt{Tang+Liu+2019}). Although a large number of analysis of variability have been studied from different perspectives, there is still no definite conclusion about what kind of process variability is.

Since the introduction of the structure function into the study of QSOs (\citealt{Simonetti+etal+1985, Hjellming+Narayan+1986, Hughes+etal+1992}), the research contents of AGNs have been further expanded. Stars and QSOs can be distinguished within the specified redshift range based on the optical variability of QSOs (\citealt{Palanque+etal+2011}), and someone has already used the structure function to distinguish stars and QSOs and detect the QSOs (\citealt{MacLeod2010, MacLeod+etal+2011, Schmidt+etal+2010, Schmidt+etal+2012, Butler+Bloom+2011, Rengstore+etal+2004}). \citet{Vries2005} used the structure function to study the variability of QSOs from SDSS, and explored its variability mechanism. In early study, \citet{Kawaguchi1998}, who aimed to infer the dominant factors about the variability through the differences between the structure function, modeled the structure function according to the variability mechanism. \citet{Hawkins2002} wanted to use structure function to directly compare with the long-term variability of a single quasar. \citet{Kozlowski2016} applied the structure function to study the variability of QSOs from SDSS, and conducted the correlation analysis based on the covariance function of the random process and the physical parameters of QSOs.

Currently, the study about the variability of AGNs has covered various bands. The study of the variability about a single band is used to explore the period of light variation and the variability mechanism, while the variability of multiple bands are used to compare the characteristics among different bands, which is a current hotspot. However, there is no specific conclusion on whether the variability characteristics of different bands can be distinguished by a certain parameter. If the variability characteristics presented in different bands are significantly different, we can suppose that it may be caused by the variability mechanism. On this basis, we can explore the physical process of variability about different bands to improve the theoretical models of the variability.

In this study, we use the structure function to analyze the variability of the radio band, optical band and gamma-ray band. To prevent the influence of the structure function values being either too big or too small on the results, we normalize the values based on the computed results (\citealt{Tang2012}). The image of structure function has two platforms, and there is an approximate linear process between these platforms (\citealt{Cristiani+etal+1996}), which represents the linear change process of the flux over time. This process can be fitted by some models such as the power function and exponential function. In this paper, we use the power function model to fit the linear process of the structure function, and obtain the power exponent representing the slope and amplitude. We distinguish the three bands by comprehensively considering these two physical quantities, then we arrange the power exponents and amplitudes of three bands in a certain order. In addition, we distinguish the RQ AGNs and RL AGNs by these two physical quantities based on the categories of the samples. Because of the use of Fermi observation data, we compare the power exponents and amplitudes of BL Lacs and FSRQs about Fermi and non-Fermi to confirm the conclusions drawn by previous works (\citealt{Linford+etal+2011, Xiong+etal+2015}).

We introduce the observation data and structure function method in sect. 2. Then, in sect. 3, we calculate the values of the structure function and fit the linear process of the samples about three bands. The discussion and conclusions are present in sect. 4.

\section{Samples and method}
\label{sect:Obs}

\subsection{Observation data}

The observation data in this paper includes 3 bands, so we need to obtain flux, magnitude and photon data of radio band, optical band and gamma-ray band.We adopted the data of radio band and classifications of the samples used in the articles by \citet{Richards2011} and \citet{Hovatta2021}. These data comes from the 40 m telescope at the Owens Valley Radio Observatory (OVRO), a unique facility in the world, and the University of Michigan Radio Observatory collaborates with OVRO. The data of radio band is dense and has a long observation time span and overall data quality. Since observations began in 2008, this observatory has monitored the flux data at 15 GHz for more than 1500 blazars and the motions of 5500 AGNs. We use the flux data of these samples at 15 GHz from 2008 to 2011. The observation data of optical band comes from the magnitudes and flux data of V-band obtained by the All-sky Automated Survey for SuperNovae (ASAS-SN) (\citealt{Christy+etal+2023}) from 2014 to 2018. The ASAS-SN is a project conducted by multiple institutions. This project primarily targets stars, and its database contains hundreds of thousands of light curves of variable and non-variable stars. We determined the classifications of these samples by searching the right ascension and declination within allowable error ranges. To consistent with other data, we use the flux data of V-band in this paper. Finally, we searched the data from NASA’s Fermi Large Area Telescope (LAT) \footnote[1]{https://fermi.gsfc.nasa.gov/} to obtain the flux data of gamma ray from 2008 to 2024. The Fermi spacecraft was launched on June 11, 2008, and the LAT which covers approximately 20 percent of the sky is the main scientific instrument for monitoring extreme gamma ray emissions. The energy range that the LAT can observe is from about 20 MeV to over 300 GeV. In this paper, we adopted flux data from the accessible samples in the 4FGL catalog and determined the classifications of these samples.

We respectively summarized the numbers of blazars and QSOs of three bands, and found that there are clear classifications about BL Lac and FSRQ in the samples of radio band and gamma-ray band. However, we can only find a few QSOs in the samples of optical band. We classified BL Lacs and FSRQs of the samples in the radio band and gamma-ray band as RL AGNs, and classified the QSOs of the samples in the optical band as RQ AGNs. We analyzed the values of the structure function and fitting results based on this classification. Before calculation, we sampled the data of three bands.

In figure 1, we show the flux curves of individual samples of three bands. It can be seen from the figure that the flux of these samples exhibits large fluctuations, indicating obvious variability.

 \begin{figure}
   \centering
   \includegraphics[width=\textwidth, angle=0]{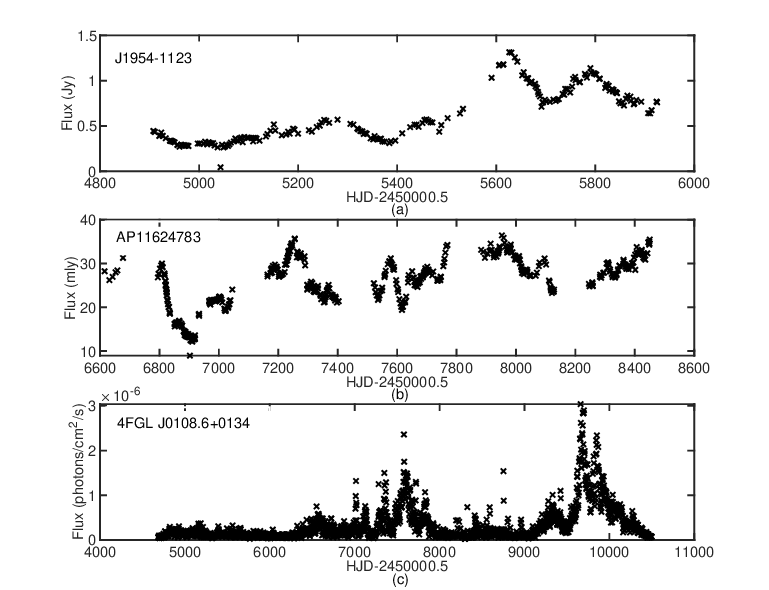}
   \caption{The light curves of the samples about three bands. (a) Radio band; (b) Optical band; (c) Gamma-ray band.}
   \label{Fig1}
   \end{figure}

\subsection{Structure function}

As a common method to study the variability of AGNs, the structure function is employed to analyze the characteristics of variability. The structure function method is non-independent. While avoiding the window problem and aliasing phenomena in Fourier analysis, it can analyze the uneven data sequences with sparse data points. We can represent the root-mean-square value of the intensity difference of variability corresponding to different time about the light curve by the structure function (\citealt{Simonetti+etal+1985}). The physical meaning of the nearly linear portion in the plot of the structure function with time delay is the rate of change about flux (\citealt{Tang2012}). The slope of this part can analyze the intrinsic variability of AGNs and QSOs, revealing the physical mechanisms of variability (\citealt{Hawkins1996, Hawkins+1993}). \citet{WangShi2020} studied the light variation characteristics of SDSS quasars in the mid-infrared band by using the structural function method.

There are three definitions of the structure function in current study. For ease of distinction, $SF^{(A)}$, $SF^{(B)}$ and $SF^{(c)}$ (\citealt{Schmidt+etal+2010, Meusinger+etal+2011, Bauer+etal+2009, Hook+etal+1994}). 
We use the $SF^{(A)}$ (\citealt{Simonetti+etal+1985, Hjellming+Narayan+1986}) that also called the first-order structure function to calculate the data. For the observed value of an object $X(t_i)$, taking any time delay $\tau$, the first-order structure function is (\citealt{Simonetti+etal+1985, Butuzova+etal+2024})
\begin{equation}\label{eq1}
  SF(\tau) = \frac{1}{N(\tau)} \sum_{i=1}^{N(\tau)} \left[ X(t_i + \tau) - X(t_i) \right]^2
\end{equation}
where $X(t_i)$ and $X(t_i+\tau)$ represent the observed value of the $i$ and $i+\tau$ points, $i = 1...N$, respectively, $N(\tau) = \Sigma [\omega (t_i)\omega (t_i+\tau)]$
represents the number of the observations. If observed values exist at the $i$ and $i+\tau$ points, the weight coefficient is $\omega (t_i) = \omega (t_i+\tau) = 1$, and if not, $\omega (t_i) = \omega (t_i+\tau) = 0$.

$SF (\tau)$ and time delay $\tau$ can be fitted well with the form of power function model (\citealt{Tang2012})
\begin{equation}\label{eq2}
  [SF(\tau)]^{\frac{1}{2}} = SF = A \tau^{\beta}
\end{equation}
where $\beta$ represents the power exponent and $A$ represents amplitude. We adopted the improved model by \citet{Tang2012} when we fitted the values of the structure function. We changed the equation of the fitting model into
\begin{equation}\label{eq3}
  SF' = A'\tau^{\beta} = (SF_{\infty} \cdot A) \tau^{\beta}
\end{equation}
\begin{equation}\label{eq4}
  SF = \frac{SF'}{SF_{\infty}} = A\tau^{\beta}
\end{equation}

We used the least square method to fit the normalized structure function into the power function model. First of all, in order to prevent the structure function is too small or too large to affect the fitting result, we normalized the structure function. The normalized structure function shows an approximately linear relationship on the double log plot of time delay $\tau$ and $SF'$, which could be fitted to a power function model (as in Eq. (4)). The value of normalized structure function and time delay were logarithmic and linear regression was performed. After fitting by the least square method, the amplitude A and the power exponent $\beta$ were obtained, both of which had certain physical significance. The amplitude reflects the change of the flow, and the power exponent reflects the rate of the flow change with time, which can reflect the mechanism of variability.

This transformation process normalized the values of the structure function, unifying the range of the values within the interval of 0 to 1. What we did about the values is to make the amplitudes more comparable without affecting the power exponent.

\section{Statistical analysis} 
\label{sect:data}

\subsection{Analysis of the structure functions}

The structure function well expresses the change of flux with time. In particular, the part between the two platforms displays the amplitude of data changes while also describing the rate of change. Because of the differences of the samples’ time spans, we prioritize data points and choose samples with more than 100 points when selecting samples to make the samples we used more representative. After preliminary,screening, we complied 923 RL AGNs of radio band, 1048 RL AGNs of gamma-ray band and 207 RQ AGNs of optical band.

In figure 2, we show the normalized structure functions and fittings of some samples in 3 bands. It can be seen that the samples of radio band, optical band and gamma-ray band all exhibit the classic structure of the structure function, which includes an approximately linear part and a platform. In figure 2 (a), we find that the overall structure function values of the sample of the radio band are in good agreement with the fitted values, with little difference between the two. For the sample of the radio band, the structure function value increases slowly after a time delay of 3 d, approximately forming a platform. The power exponent of the fitting function for this sample is 0.65, and a linear part appears in the time delay interval of 1 to 3 d, where the power exponent is approximately equal to the slope. In figure 2 (b), we similarly observe that the structure function values of the sample of optical are also in good agreement with the fitted values, with the small differences between the two. For the sample of the optical band, the structure function value increases slowly after a time delay of 3.5 d, forming a platform. In the time delay interval 1.5 to 3.5 d, the power exponent of the fitting function is approximately equal to the slope of the linear part at 0.72. In figure 2 (c), we find that the differences between the structure function values and the fitted values of the sample of gamma ray is large, indicating poor fitting performance. For this sample, a platform approximately appears after a time delay of 10 d, and the structure function shows an linear part in the time delay interval of 2 to 10 d, where the slope is close to the power exponent of the fitting function at 0.40. Through the analysis of the structure functions and fittings of the samples of 3 bands, it can be seen that after normalizing the values of the structure function, the two parameters obtained from the fitting both have physical meanings.

Through the uncertainty (1$\sigma$ confidence interval) in Figure 2, we find that the prediction fluctuation of the model in the whole time delay range is controlled in the sample fitting process of the optical and radio bands, indicating that the power function model has a strong constraint ability on the data. Even within the large delay range, the predictive reliability of the samples of gamma-ray band was not affected. It can be seen from Figure 2 that the samples in the optical and radio bands deviate less from the residual zero line, while the residual of gamma-ray band is larger. This indicates that the power function model can describe the evolution process of structure function of different types of sources under different time scales,and has a better fitting effect on the optical and radio bands, while the fitting effect of gamma-ray band is relatively poor.

For the structure function of gamma-ray, the power function model fits poorly. There are two possible reasons. Compared with the radio (OVRO) and optical (ASAS-SN) surveys, the noise of the Fermi Large Area Telescopes’s gamma-ray observations is higher. Physically, the radiation of high-energy gamma-ray is usually associated with extremely relativistic jet, which presents a more complex pattern of variability than radio and optical radiation. Large measurement errors and complex pattern of variability lead to additional discreteness in the structure function, thereby reducing the goodness of fit. \citet{Giacchinoetal2024} argued that the variability of gamma-ray of blazars are usually dominated by nonlinear physical processes, which cannot be completely captured by simple power function models.

 \begin{figure}
   \centering
   \includegraphics[width=\textwidth, angle=0]{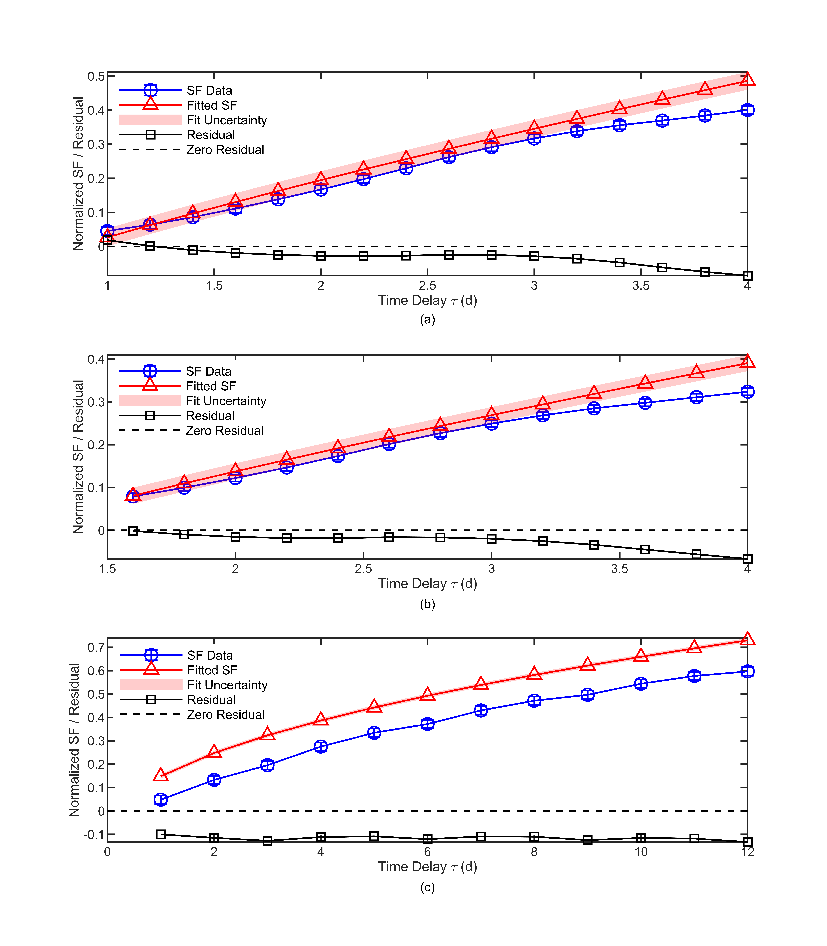}
   \caption{The structure functions of the samples about three bands. (a) Radio band; (b) Optical band; (c) Gamma-ray band.}
   \label{Fig2}
   \end{figure}

\subsection{Parameter statistical analysis}

We counted the amplitudes and power exponents of 3 bands respectively and plotted histograms of these two parameters after calculating the structure functions for all samples of 3 bands and fitting the values of the structure functions. The distribution intervals of the amplitudes and power exponents of the samples about 3 bands are all shown in figure 3. From the amplitude distribution histogram in Figure 3 (a) (the solid line represents the radio band, the dashed line represents the gamma-ray band and the dotted line represents the optical band). The amplitudes of the samples about 3 bands overlap in the range of 0.28 - 0.42, but the peak ranges of each band are different. From the power exponent distribution histogram in figure 3 (b), it can be seen that a small amount of overlap in the power exponents of the samples about 3 bands, but the peak ranges of each band show obvious differences. By analyzing the amplitude and power exponent distribution histograms of the samples about 3 bands comprehensively, we found that although the amplitudes are not easy to distinguish the samples of the optical band and the radio band, the power exponents can completely distinguish the samples about 3 bands. By integrating the amplitude and power exponent distribution of 3 bands, we sorted their power exponents, and the order obtained according to the statistical results is that the optical band is the largest, followed by the radio band, and finally gamma-ray band. This conclusion can be used as a basis for analyzing  the variability mechanism and can also be used to improve the models of variability.

 \begin{figure}
   \centering
   \includegraphics[width=\textwidth, angle=0]{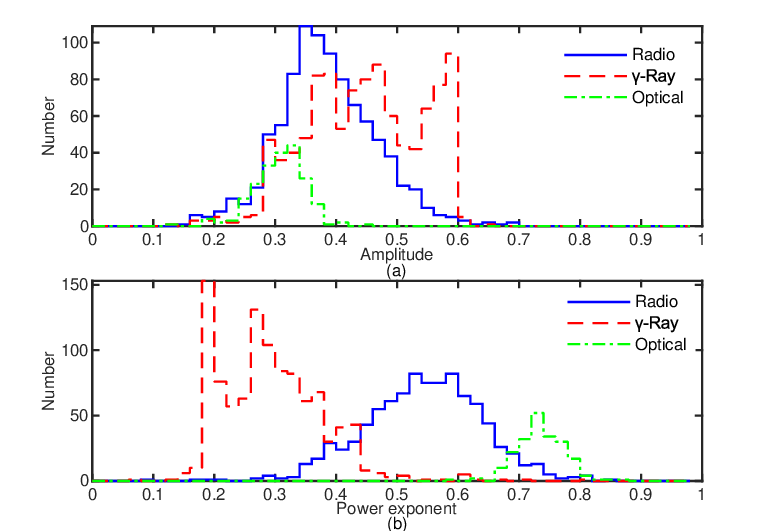}
   \caption{The distribution of the fitted parameters. (a) Histogram of amplitude distribution; (b) Histogram of power exponent distribution.}
   \label{Fig3}
   \end{figure}

Based on the amplitude and power exponent distribution histograms, we plotted the amplitude-power exponent scatter plots for samples of 3 bands. To quantitatively delineate the parameter distribution characteristics of different sample categories, we employ a statistical critical sampling method to partition the corresponding regional boundaries within the parameter plane. Firstly, the contour point sets of each category are identified within the parameter space, which correspond to the outermost data points of the respective distributions. Subsequently, linear fitting is performed separately on the contour points of each category to derive linear boundaries that discriminate the distribution ranges of different sample categories represented by solid and dashed black lines in Figure 4. Through this boundary delineation method based on contour point statistic and linear fitting, the distribution scopes of the three sample categories can be clearly defined in the two-dimensional “amplitude-power exponent” space.

As shown in Figure 4, there are distinct boundaries between the parameters of the samples about 3 bands, which are distributed in three different regions. The samples of gamma-ray are characterized by high amplitude and low power exponent distributions. Optical-band samples are predominantly concentrated in regions of low amplitude and high power exponent. In contrast, the samples of radio band exhibit an intermediate distribution in the parameter space, displaying a transitional morphology between the aforementioned two categories. The parameter distribution regions of each band's samples in Figure 4 correspond one-to-one with the amplitude distribution histogram and power exponent distribution histogram in Figure 3. We divided the three regions with two lines. Obviously, the points of three different shapes are respectively clustered within three ranges. Meanwhile, according to our classification of samples from different bands, we can also distinguish between RQ AGNs and RL AGNs. The samples selected for the radio band and gamma-ray band are RL AGNs, while those selected for the optical band are RQ AGNs. Although there is significant overlap in amplitude, the overlapping portion of power exponents is minimal. The right part of figure 4 represents RQ AGNs, and the left and middle parts represent RL AGNs, with a clear boundary between the two regions. Therefore, based on the two fitted parameters, we can not only distinguish between RQ AGNs and RL AGNs but, more importantly, differentiate the samples of 3 bands.

 \begin{figure}
   \centering
   \includegraphics[width=\textwidth, angle=0]{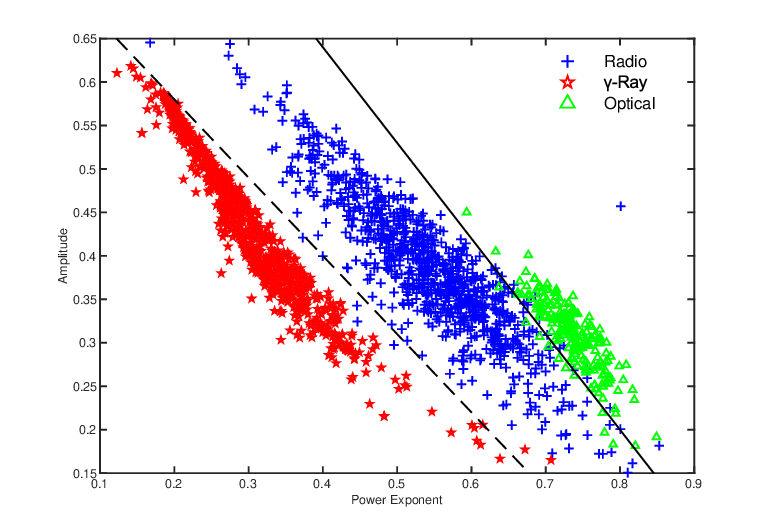}
   \caption{Amplitude-power exponent scatter plot. In the figure, the triangular markers represent samples of the optical band, the cross markers represent samples of the radio band, and the five-pointed star markers represent samples of the gamma-ray.}
   \label{Fig4}
   \end{figure}

To more intuitively understand the distribution of amplitude and power exponent, we statistically analyzed the number of occurrences of parameters for samples of three bands within different intervals and their proportions in the total number of samples. The specific statistical results are presented in Table 1 and Table 2. According to the statistical results in Table 1, the amplitudes of samples of each band are concentrated in specific intervals. The amplitudes of radio-band samples are mainly concentrated in two intervals, optical-band samples in two intervals, and gamma-ray-band samples in two intervals. From the statistical results in Table 2, the power exponents of each band's samples also have obvious concentrated intervals: radio-band samples are most concentrated in one interval, optical-band samples are mainly concentrated in one interval, and gamma-ray-band samples are concentrated in one interval. Through the analysis of Table 1 and Table 2, we can not only intuitively understand the concentrated regions of sample parameters in each band but also find that the statistical data in the tables correspond to the histogram in figure 3 and the scatter plot in figure 4. Therefore, using these two parameters, we can not only distinguish samples of the radio band, optical band and gamma-ray band but also differentiate between RQ AGNs and RL AGNs based on sample classification.

%

\begin{table}
\centering
\caption[]{Amplitude Statistics Table}\label{Tab1}
\label{tab:amplitude_stats}
\setlength{\tabcolsep}{3pt} 
\small 
\begin{tabular}{l *{10}{c}}
\hline
 & \multicolumn{2}{c}{0.20--0.30} & \multicolumn{2}{c}{0.30--0.40} & \multicolumn{2}{c}{0.40--0.50} & \multicolumn{2}{c}{0.50--0.60} & \multicolumn{2}{c}{0.60--0.70} \\
\cline{2-11}
 & Number & Percent (\%) & Number & Percent (\%) & Number & Percent (\%) & Number & Percent (\%) & Number & Percent (\%) \\
\hline
Radio & 106 & 11.48 & 445 & 48.21 & 288 & 31.20 & 63 & 6.83 & 9 & 0.98 \\
Optical & 76 & 36.71 & 123 & 59.42 & 3 & 1.45 & 0 & 0.00 & 0 & 0.00 \\
Gamma & 65 & 6.20 & 289 & 27.58 & 355 & 33.87 & 321 & 30.63 & 6 & 0.57 \\
\hline
\end{tabular}
\end{table}

\begin{table}
\caption[]{Power Index Statistics Table}\label{Tab2}
\label{tab:amplitude_stats}
\setlength{\tabcolsep}{3pt} 
\small 
\begin{tabular}{l *{10}{c}}
\hline
 & \multicolumn{2}{c}{0.20--0.30} & \multicolumn{2}{c}{0.30--0.40} & \multicolumn{2}{c}{0.40--0.50} & \multicolumn{2}{c}{0.50--0.70} & \multicolumn{2}{c}{0.70--0.90} \\
\cline{2-11}
 & Number & Percent (\%) & Number & Percent (\%) & Number & Percent (\%) & Number & Percent (\%) & Number & Percent (\%) \\
\hline
Radio & 7 & 0.76 & 64 & 6.93 & 213 & 23.08 & 596 & 64.57 & 40 & 4.33 \\
Optical & 0 & 0.00 & 0 & 0.00 & 0 & 0.00 & 35 & 16.91 & 172 & 83.09 \\
Gamma & 431 & 41.13 & 325 & 31.01 & 101 & 9.64 & 13 & 1.24 & 2 & 0.19 \\
\hline
\end{tabular}
\end{table}

Since the samples in the radio and gamma-ray have clear classifications of BL Lac and FSRQ, and the gamma-ray sample data are from the Fermi telescope. Some scholars have currently studied blazars observed by the Fermi telescope and non-Fermi telescopes, concluding that their BL Lacs are essentially indistinguishable, but FSRQs show significant differences. We compared the parameters of BL Lac and FSRQ samples obtained from radio telescopes with those of BL Lac and FSRQ obtained from the Fermi telescope to verify whether they align with previous conclusions. In figure 5, we show the amplitude and power exponent distributions of BL Lac samples from radio telescopes and the Fermi telescope. As seen in Figure 5(a), the amplitudes of BL Lac samples of Fermi and non-Fermi have overlapping distribution ranges and consistent peak positions. However, Figure 5(b) reveals a clear boundary in the power exponents of the two BL Lac samples, with peaks appearing in different intervals. In terms of power exponents, only a small portion overlaps, and the peak regions differ significantly. From figure 6(a), we observe that the amplitudes of the Fermi and non-Fermi FSRQ samples have overlapping ranges but also show differences in peak distribution. Therefore, based on the peak ranges of the amplitudes of the Fermi and non-Fermi FSRQ samples, the two sets of FSRQ samples can be distinguished. According to figure 6(b), we clearly observe that the peaks of the power exponents of the Fermi and non-Fermi FSRQ samples are in two different regions with a clear boundary. The conclusions drawn from the analysis of figures 5 and 6 are largely consistent with those of previous studies. However, in the analysis of their BL Lac samples, we found that the power exponent can be used for differentiation, which is somewhat different from the conclusions of previous studies. Considering the number of BL Lac samples from the two telescopes, we suggest that a possible reason is the difference in sample size: there are 476 BL Lac samples from the Fermi telescope, while only 111 from the radio telescope, meaning the number of Fermi BL Lac samples is more than four times that of non-Fermi BL Lac samples. The huge disparity in sample size may lead to uneven distributions of the power exponents. Nevertheless, the conclusions from the analysis of the FSRQ samples are consistent with the results of previous studies.

 \begin{figure}
   \centering
   \includegraphics[width=\textwidth, angle=0]{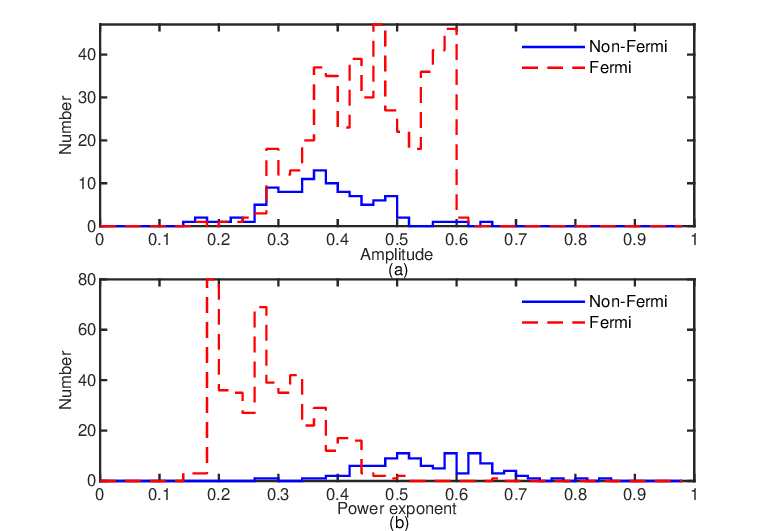}
   \caption{Parameter distributions of Fermi and non-Fermi BL Lacs. (a) Histogram of amplitude distribution; (b)Histogram of power exponent distribution.}
   \label{Fig5}
   \end{figure}

\begin{figure}
   \centering
   \includegraphics[width=\textwidth, angle=0]{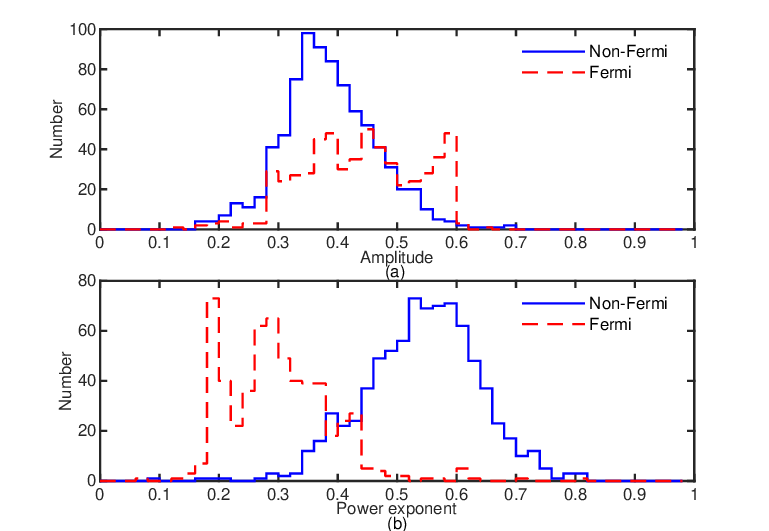}
   \caption{Parameter distributions of Fermi and non-Fermi FSRQs. (a) Histogram of amplitude distribution; (b) Histogram of power exponent distribution.}
   \label{Fig6}
   \end{figure}
   
\section{Discussion and conclusions}
\label{sect:analysis}
There are three theoretical models explaining the variability mechanisms of QSOs and AGNs mentioned in this work, namely the disk instability model, the starburst model, and the gravitational lensing model. The first two models suggest that variability are caused by intrinsic changes in celestial objects, while the last model attributes them to external environmental factors (\citealt{Hawkins+1993}). However, regardless of the theoretical model, each can explain the variability of QSOs and AGNs observed in practice. Most scholars believe that supermassive black holes exist in the central regions of QSOs and AGNs, and the thermal radiation from the central black hole accretion disk may be one of the causes of long-term variability in these celestial objects (\citealt{Zhang+2004, Deng+Huang+2008}).

\citet{Hawkins1996} simulated the variability of QSOs based on these three theoretical models and derived the common characteristics of their structure functions. The results of linear power-law fitting of structure functions differ across models. The gravitational lensing model yields the smallest fitting slope, ranging from 0.22 to 0.28; the starburst model gives the largest slope, ranging from 0.74 to 0.90; and the disk instability model falls between them, with a slope range of 0.41 to 0.49. In the samples selected in this study, most power exponents fitted from radio-band samples are distributed between 0.40 and 0.70, making the disk instability model suitable for explaining the variability of AGNs. The optical-band samples show a large concentration of power exponents between 0.70 and 0.90, which is consistent with the starburst model. Gamma-ray samples, however, have power exponents mainly distributed between 0.20 and 0.30, supporting the gravitational lensing model for their variability. This finding implies that the variability of a large proportion of radio and optical bands are mainly caused by intrinsic changes in celestial objects, while the variability of gamma-ray band is driven by external environmental factors. In this paper, the power exponent is not the only criterion to constrain the physical model, but rather a quantitative indicator related to the time characteristics of variability, which can be used to correlate with physical process. When the slope is steep ($\beta$ ranges from 0.70 to 0.90), the variability changes rapidly with time delay, which is consistent with the starburst model. The moderate slope ($\beta$ range around 0.40) matches the disk instability model, the random characteristics of accretion disk fluctuation leads to a slow increase of variability with time delay. When the slope is flat (with a $\beta$ range of 0.20 to 0.30), it is more consistent with gravitational lensing.

The range of power exponents of gamma-ray samples is basically consistent with the simulation results of the gravitational lensing model by \citet{Hawkins1996}, and recent study have also shown that the jet radiation of some gamma-ray blazars is modulated by the lensing model (\citealt{Giacchinoetal2024}). The gravitational lensing model suggests that this slowing of the slope of the structure function is due to the gradual and smooth amplification of the dense object in the middle. In contrast, the intrinsic jet process involves rapid energy dissipation and more intense variability (\citealt{Giacchinoetal2024}; \citealt{Amador+etal+2024}; \citealt{Zackrisson+etal+2003}). The micro-gravitational lensing of gravitational lensing is the core cause of the short-term variability of gamma-ray band. The gamma-ray band's emission from AGN are produced by relativistic jets, and their emission region is extremely dense (\citealt{Giacchinoetal2024}). According to general relativity, the gravitational field of a compact object will bend the path of electromagnetic radiation propagation. Only when the lensing body, the AGN jet, and the observer are almost in a straight line, will the deflection effect be significantly enhanced (\citealt{Einstein+1936}; \citealt{Blandford+Narayan+1992}). For the long-term variability, they are caused by macro-gravitational lensing. This process indicates that the large-scale mass distribution of galaxies will cause the gamma-ray band's photons of AGN to be deflected along multiple different paths of propagation, resulting in the flux of gamma-ray band received by the observer being the superposition of multiple delayed image fluxes. Small-scale fluctuations in a single image will cause the superimposed flux to exhibit slow long-term fluctuations. Under this process, the variability timescale is usually several years, and the flux fluctuations are smooth without sharp mutations, leading to an overall low amplitude of the light curve of gamma-ray band (\citealt{Paynter+etal+2021}; \citealt{Oguri+Kawano+2003}; \citealt{Treu+2010}). 

Based on the samples we have obtained, in their light curves of long-term, the amplitude of variability is not significant, and for the more obvious samples, the amplitude fluctuation time is relatively short, which conforms to the variability within a short time scale (\citealt{Jaron+etal+2024}). In the radio and optical bands, due to various emission mechanisms and other reasons, the processes related to the jet effects may dominate. Gamma-ray band, on the other hand, usually originates from inverse compton scattering in compact regions and may be more susceptible to external modulation influences (\citealt{La+etal+2017}; \citealt{Bai+Lee+2003}; \citealt{Dubus+etal+2008}; \citealt{Mondal+Gupta+2019}).

Therefore, although jet activity remains the main driver of the high-energy radiation in quasars, its influence on the structure function may be masked by external factors in the gamma-ray band (such as gravitational lensing effects), especially in the time scale and sample composition considered in this study. This explanation is consistent with the latest findings of \citet{Giacchinoetal2024} and \citet{Wagneretal2025}, who also reported evidence of external modulation in gamma-ray quasars.

Based on these studies and the statistical analysis in this paper, 72.14\% of the samples of gamma-ray band in this study fall within the predicted range of the gravitational lensing model, indicating that external factors are the main driving force in the samples of gamma-ray band we have studied. We emphasize that although most of the gamma-ray band's samples in this paper are modulated by the gravitational lensing model, this does not negate the role of jets. Jets, as the radiation sources of gamma-ray band are also crucial, and the gravitational lens is the modulator of the variability of the samples of gamma-ray band in this paper. The short-term flux amplification effect of the micro-gravitational lens and the long-term multi-image superposition effect of the macro-gravitational lens become the dominant external factors for the variability of gamma-ray band in this paper (\citealt{Giacchinoetal2024}).

\citet{MacLeod2012} proposed that the slope of structure function around 0.5 aligns with predictions from the damped random walk model. Given that both the starburst and disk instability models support the idea that the variability arise from random events, our results indicate that the variability of radio band is more consistent with the damped random walk model.

Since the sample types of the optical band are RQ AGNs (mainly QSOs), while the sample types of radio and gamma rays are RL AGNs (including BL Lacs and FSRQs). It can be seen from Figure 4 that there is a clear dividing line between RQ AGNs and RL AGNs, indicating significant differences between these two types of AGN. This conforms to the unified standard explanation of AGN in recent years. The variability of RQ AGNs is mainly caused by the thermal radiation fluctuations of the accretion disk (\citealt{WuLi1996}), while the variability of RL AGNs is caused by the relativistic jet radiation enhanced by the Doppler effect (\citealt{Jaron+etal+2024, Paliya+2024}). The jet process is closely related to the accretion process (\citealt{Chen+etal+2022}). At the same time, it also confirms the differences in variability mechanisms of the high-energy band. That is, in the high-energy radio and gamma ray fields, due to relativistic jets, the phenomena of variability can be more easily observed, especially for Blazars (\citealt{Giacchinoetal2024}).

For the observational data from Fermi and non-Fermi telescopes, researchers (\citealt{Linford+etal+2011, Xiong+etal+2015}) argue that there is no fundamental difference between their BL Lac objects, but significant differences exist in FSRQs. We further expanded the sample sizes of Fermi telescope and radio telescope observations, and verified this conclusion using the amplitudes and power exponents derived from fitting structure functions. The results show that Fermi and non-Fermi BL Lac samples exhibit no obvious separation in amplitudes, but significant partitioning in power exponents. We suggest that the large discrepancy in sample sizes between the two BL Lac datasets may cause their power exponents to concentrate in different intervals, leading to conclusions different from previous studies. However, we perfectly validated the conclusion that Fermi and non-Fermi FSRQs are significantly different. Both amplitude and power exponent show clear interval boundaries between the two, indicating significant differences. Through our analysis, we can distinguish Fermi and non-Fermi blazars based on these two parameters, confirming previous conclusions while also deriving new insights from the analytical results.

In this study, we introduced amplitude based on the power exponent. Although the amplitude cannot distinguish the samples from the three bands in the histogram, the power exponent perfectly discriminates the samples across the three bands. The purpose of introducing amplitude lies in the division of sample regions in the scatter plot. When the two parameters are plotted in the same graph, the parameters of the samples from the three bands are distributed in three distinct regions, with clear boundaries between the regions. Additionally, using amplitude, we confirmed that there is no fundamental difference between Fermi and non-Fermi BL Lacs. From the amplitude distribution results alone, we cannot distinguish the BL Lacs samples between the two, but they can be distinguished by the power exponent, and we also analyzed the reasons for this. Based on these two parameters, we validated the conclusions of previous studies.

The conclusions drawn in this paper are of great significance for distinguishing the samples of different bands and provide a basis for differentiating between RQ AGNs and RL AGNs. Some current studies have conducted correlation analyses on variability in different bands (\citealt{Li+etal+2020, Zhang+etal+2021}) to explore the similarities and differences in variability mechanisms across bands. The conclusions of this paper can also offer certain support for analyzing the detailed principles of variability phenomena in different bands through structure functions.

\begin{acknowledgements}
This work utilized data from the OVRO 40 m Monitoring Program (\citealt{Richards2011, Hovatta2021}), which was supported by NSF grants AST-0808050 and AST-1109911, and is currently supported by NSF grants AST-2407603 and AST-2407604. It was also funded by NASA NNX08AW31G, NNX11A043G, and NNX14AQ89G. We acknowledge the gamma-ray data released on the Fermi website by NASA, and we thank the ASAS-SN project for providing data (\citealt{Christy+etal+2023}). We appreciate the hard work of all project researchers. Meanwhile, we are grateful for the support of the National Natural Science Foundation of China (11373008) and the Natural Science Foundation of Shaanxi Province (2013JM1021).
\end{acknowledgements}
\section*{Data Availability}
The data used in this study are publicly available in the OVRO, ASAS-SN DR9, and Fermi-4FGL Data Release.

\label{lastpage}

\end{document}